\documentclass[a4paper,10pt,twoside]{cpc-hepnp}
\usepackage{upgreek,fancyhdr}
\usepackage{multicol}
\usepackage{graphicx}
\usepackage{booktabs}
\usepackage{amssymb,bm,mathrsfs,bbm,amscd}
\usepackage[tbtags]{amsmath}
\usepackage{lastpage}
\usepackage{color}

\begin{document}

\fancyhead[c]{\small Submitted to Chinese Physics C}
\fancyfoot[C]{\small 010201-\thepage}

\footnotetext[0]{\hspace*{-3mm}\raisebox{0.3ex}{$\scriptstyle\copyright$}This work was supported in part by
the Ministry of Science and Technology of the People’s Republic of
China (2015CB856703), by by the Strategic Priority Research Program
of the Chinese Academy of Sciences (XDB23030000), and by the National
Natural Science Foundation of China(NSFC) under the grant 11375200.}%

\title{X(16.7) as the Solution of NuTeV Anomaly}

\author{%
      Yi Liang $^{1;1)}$\email{alavan@ucas.ac.cn}%
\quad Long-Bin Chen $^{1,2;2)}$\email{chenlongbin@ucas.ac.cn}%
\quad Cong-Feng Qiao $^{1,2;3)}$\email{qiaocf@ucas.ac.cn}
}
\maketitle

\address{%
$^1$ School of Physics, University of Chinese Academy of Sciences, YuQuan Road 19A, Beijing 100049, China\\
$^2$ CAS Center for Excellence in Particle Physics, Beijing 100049, China\\
}

\begin{abstract}
A recent experimental study of excited $^{8}\!Be$ decay
to its ground state reveals an anomaly in the final states angle distribution. This exceptional result is attributed to a new vector gauge boson X(16.7). We study the significance of this new boson, especially its effect in anomalies observed in long-lasting experimental measurements. By comparing the discrepancies between the standard model predictions and the experimental results, we manage to find out the values and regions of the couplings of X(16.7) to muon and muon neutrino. In this work, we find that the newly observed boson X(16.7) may be the solution of both NuTeV anomaly and the $\left(g-2\right)_{\mu}$ puzzle.
\end{abstract}



\begin{multicols}{2}

\section{Introduction}

As a theory describing electroweak and strong interactions, the Standard
Model (SM) has achieved great success, and has been tested at high
precision. However, some experimental studies pointed out the possibility
for new physics beyond SM. For example, the non-zero masses of neutrinos,
the existence of dark matter, the muon anomalous magnetic moment etc..
More fundamental challenges such as the hierarchy problem also put
severe challenges for the Standard Model in describing the nature.
Searching for new physics beyond the Standard Model (BSM) has become
one of the major activities in physics. Numerous new physics models
have been proposed. One of the simplest possibilities is $SU(3)\times SU(2)_{L}\times U(1)$
extended by a new gauge group $U(1)$.

A result in $^{8}\!Be$ nuclear transition brought forth new challenges
to our understanding of electroweak interaction. In this reaction,
$^{8}\!Be$ decays from an excited state to its ground state $^{8}\!Be^{*}$
$\rightarrow$ $^{8}\!Be\ X$, followed by saturating decay $X\to e^{+}e^{-}$.
A $6.8\sigma$ anomaly to the internal pair production was observed
at a angle of $140^{\circ}$ \cite{Krasznahorkay:2015iga}. Although
this extraordinary experimental phenomenon may due to unidentified
nuclear reactions or experimental errors, it can also be attributed
to a new vector boson X with mass of 16.7 MeV, which mediates a weak
fifth force BSM. In other words, the SM gauge group is extended by
a new Abelian gauge groups $U(1)_{X}$, which is one of the most natural
extension of the SM \cite{Feng:2016jff}. Based on this hypothesis,
the values and regions of the first-generation charges of this protophobic
gauge boson was investigated. New renormalizable model for this vector
boson is proposed \cite{Gu:2016ege}. The possibility of revealing
this yet to be verified gauge boson in other electron-positron colliders,
such as at BESIII and BaBar is evaluated \cite{Chen:2016dhm}.

Other than this phenomenal experimental discovery, discrepancies between
experiment data and SM predictions were exposed by several relatively
old experimental studies, such as the anomalous magnetic moment of
muon $g-2$, and the NuTeV anomaly \cite{Zeller:2001hh}. NuTeV experiment found a $3\sigma$
deviations above the SM prediction for $\sin^{2}\theta_{W}$, large
discrepancy between theoretical calculation and experimental measurements
were also found earlier in an experiment measuring the $\bar{\nu}_{e}e$
elastic scattering cross-section \cite{Reines:1976pv}. Other experiments
seems point to the same direction, that there is contribution from
new physics in electroweak interactions \cite{Vilain:1993kd,Vilain:1994qy,Casalbuoni:1999mw,Bennett:2004pv}.
It is pointed out that the existence of a new light gauge boson seem
to be one of the most natural explanations \cite{Altmannshofer:2014pba},
especially the muon $g-2$ anomaly can be related to a light vector
boson $Z_{\mu}$ \cite{Gninenko:2014pea}. It is tempting to see whether
the gauge boson X is responsible for these experimental anomalies.
In this work, we study BSM effect introduced by the Abelian gauge
boson X to several well known experimental results, and investigate
the values and regions of the coupling constant of this protophobic
fifth force mediator to muon and muon neutrino especially.

\section{NuTeV anomaly}

As mentioned before, the discrepancy found by NuTeV experiment is
a well known result. It has been discussed by many literatures. The
explanations of the three standard deviation above the SM predictions
for the value of $\sin^{2}\theta_{W}$ may come from the SM or BSM
have been proposed \cite{Davidson:2001ji,McFarland:2003jw,Kretzer:2003wy,Olness:2003wz}.
However, no definite conclusion can be made due to large uncertainties.
In this work, we show that the NuTeV anomaly can be fully attributed
to the contribution of the X boson. The corresponding couplings of
this new gauge boson can be chosen without contradicting to the constrains
given in \cite{Feng:2016jff}.

We use the same Lagrangian proposed in \cite{Feng:2016jff}. The $16.7\:\mathrm{MeV}$
Abelian gauge boson X with field strength tensor $X_{\mu\nu}$ couples
non-chirally to the SM fermions through vector current $L=-\frac{1}{4}X_{\mu\nu}X^{\mu\nu}+\frac{1}{2}m_{X}^{2}X_{\mu}X^{\mu}-JX^{\mu}$.
The corresponding charge is noted as $\varepsilon_{f}$ in unites
of $e$. The current $J_{\mu}=\sum_{f}e\varepsilon_{f}\,\bar{f}\gamma_{\mu}f$
, however, can still be split into left-handed and right-handed pieces
$J_{\mu}=\sum_{f}e\varepsilon_{f}\,\bar{f}_{L}\gamma_{\mu}f_{L}+\sum_{f}e\varepsilon_{f}\,\bar{f}_{R}\gamma_{\mu}f_{R}$.
According to this model, apparently, the left-handed and right-handed
fermions have identical charge. The mass of the X boson is far smaller
than the center of mass energy of major electron-positron colliders.
We adopt the conclusion given in \cite{Feng:2016jff}, that the charges
for up and down quarks satisfy the relation $\varepsilon_{d}=-2\varepsilon_{u}$.
On the other hand, as illustrated in Ref. \cite{Feng:02}, if isospin is conserved for the decay studied in the Atomki experiment \cite{Krasznahorkay:2015iga}, the summation of $\varepsilon_{u}$
and $\varepsilon_{d}$ is constrained by
\begin{equation}
|\varepsilon_{u}+\varepsilon_{d}|\approx\frac{3.3\times10^{-3}}{\sqrt{\mathrm{Br}\left(\chi\to e^{+}e^{-}\right)}}\label{eq:Feng's charges}
\end{equation}
 Notice in this charge assignment, quark universality has been relaxed.
The upper bound on $|\varepsilon_{e}|$ is provided by the measurement of electron magnetic moment $(g-2)_e$ \cite{Davoudiasl:2014kua}. The lower bound on $|\varepsilon_{e}|$ is given by the SLAC experiment E141 \cite{Riordan:1987aw,Bjorken:2009mm}. The most strict upper bound on the coupling between electron and electron neutrino comes from the TEX-ONO experiment in Taiwan \cite{Deniz:2009mu}. These constraints can be summarized as follows
\begin{align}
 & 2\times10^{-4}\le|\varepsilon_{e}|\le1.4\times10^{-3},\nonumber \\
 & |\varepsilon_{\nu_{e}}\varepsilon_{e}|^{1/2}\le7\times10^{-5}.\label{eq:Feng constraints1}
\end{align}
The $\left(g-2\right)_{\mu}$ puzzle can be solved with $\varepsilon_{\mu}$
falling in the same range as $\varepsilon_{e}$. We will find out
the constraint on $\varepsilon_{\mu}$ coming form the results of
NuTeV, and the effect introduced by particle X to the number of neutrino
flavors.

First of all, let us look at the effective four-fermions Lagrangian
generated by X exchange given in \cite{Feng:2016jff}
\begin{align}
 & L_{X}=-\frac{e^{2}}{2\left(m_{X}^{2}-t\right)}\bigg[\varepsilon_{u}\bar{u}_{L}\gamma_{\mu}u_{L}+\varepsilon_{d}\bar{d}_{L}\gamma_{\mu}d_{L}+\varepsilon_{u}\bar{u}_{R}^{c}\gamma_{\mu}u_{R}^{c}\nonumber \\
 & +\varepsilon_{d}\bar{d}_{R}^{c}\gamma_{\mu}d_{R}^{c}+\varepsilon_{\nu_{\mu}}\bar{\nu}_{\mu}\gamma_{\mu}\nu_{\mu}+\ldots\bigg]^{2}\label{eq:LagX}
\end{align}
In NuTeV experiment, nucleon are scattered by $\nu_{\mu}$, the corresponding
effective Lagrangian in SM at tree level can be expressed as
\begin{align}
 & L_{eff}=-2\sqrt{2}G_{F}\left(\left[\bar{\nu}_{\mu}\gamma_{\alpha}\mu_{L}\right]\left[\bar{d}_{L}\gamma^{\alpha}u_{L}\right]+h.c.\right)\nonumber \\
 & -2\sqrt{2}G_{F}\sum_{A,q}g_{Aq}\left[\bar{\nu}_{\mu}\gamma_{\alpha}\nu_{\mu}\right]\left[\bar{q}_{A}\gamma^{\alpha}q_{A}\right],\label{eq:Lageff}
\end{align}
where $A=\left\{ L,R\right\} $, $q=\left\{ u,d,s,\ldots\right\} $
and the couplings $g_{Aq}$ are in terms of the weak mixing angle
$s_{W}\equiv\sin\theta_{W}$. The transfer momentum squared adopted
by NuTeV is $t=-Q^{2}=-20\,\mathrm{GeV^{2}}$. What NuTeV measured
is the ratio of neutral-current to charged-current deep-ineleastic
neutrino-nucleon scattering total cross-sections. In standard model
this ratio is given by
\begin{align}
 & R=\frac{\mathrm{neutral\:currents}}{\mathrm{charged\:currents}}=\frac{\sigma_{\left(\nu_{\mu}N\to\nu_{\mu}X\right)}-\sigma_{\left(\bar{\nu}_{\mu}N\to\bar{\nu}_{\mu}X\right)}}{\sigma_{\left(\nu_{\mu}N\to\mu X\right)}-\sigma_{\left(\bar{\nu}_{\mu}N\to\mu^{+}X\right)}}\nonumber \\
 & =\left(g_{l}^{2}-g_{r}^{2}\right)=\frac{1}{2}-\sin^{2}\theta_{W}
\end{align}
where $g_{l}^{2}\equiv g_{Lu}^{2}+g_{Ld}^{2}=\frac{1}{2}-\sin^{2}\theta_{W}+\frac{5}{9}\sin^{4}\theta_{W}$,
and $g_{R}^{2}\equiv\frac{5}{9}\sin^{4}\theta_{W}$. The standard
model prediction with parameters determined by a fit to electroweak
measurements is $\sin^{2}\theta_{W}=0.2227\pm0.0004$ \cite{Abbaneo:2001ix},
while the NuTeV result is $3\sigma$ higher $\sin^{2}\theta_{W}^{\left(on-shell\right)}=0.2277\pm0.0013$.
We next find out how the value of $\sin^{2}\theta_{W}$ is altered
by the new gauge boson X, by calculating the effects of X boson to
the coupling constants $g_{L}$ and $g_{R}$. Comparing (\ref{eq:LagX})
and (\ref{eq:Lageff}), we obtain the contributions of the X mediated
tree level process to the coupling constants
\begin{align}
 & \delta g_{Lu}=\frac{\varepsilon_{u}\varepsilon_{\nu_{\mu}}e^{2}}{2\sqrt{2}G_{F}\left(m_{X}^{2}+Q^{2}\right)}\nonumber \\
 & \delta g_{Ld}=\frac{-2\varepsilon_{u}\varepsilon_{\nu_{\mu}}e^{2}}{2\sqrt{2}G_{F}\left(m_{X}^{2}+Q^{2}\right)}\nonumber \\
 & \delta g_{Ru}=\frac{\varepsilon_{u}\varepsilon_{\nu_{\mu}}e^{2}}{2\sqrt{2}G_{F}\left(m_{X}^{2}+Q^{2}\right)}\nonumber \\
 & \delta g_{Rd}=\frac{-2\varepsilon_{u}\varepsilon_{\nu_{\mu}}e^{2}}{2\sqrt{2}G_{F}\left(m_{X}^{2}+Q^{2}\right)}
\end{align}
Accordingly, the modification of $\sin^{2}\theta_{W}$ is $\delta\sin^{2}\theta_{W}=-\delta\left(g_{L}^{2}-g_{R}^{2}\right)=\frac{6\pi\alpha\varepsilon_{u}\varepsilon_{\nu_{\mu}}}{\sqrt{2}G_{F}\left(m_{X}^{2}+Q^{2}\right)}\approx5\times10^{-3}$.
Assuming $\varepsilon_{\nu_{\text{\ensuremath{\tau}}}}\sim\varepsilon_{\nu_{\mu}}$,
we obtain the charges
\begin{align}
 & \varepsilon_{\nu_{\mu}}\simeq\pm2.0\times10^{-3}\label{eq:numucharge}\\
 & \varepsilon_{u}\simeq\pm5.7\times10^{-3}
\end{align}
by combining this formula with (\ref{eq:Feng's charges}) and taking
the upper limit of $\varepsilon_{e}\sim1.4\times10^{-3}$. The difference
between the experimental value and standard model expectation of the
Weinberg angle is resolved. It is worth noting that if the NuTeV anomaly
is entirely due to the new $\mathrm{U}\left(1\right)$ particle X
, the absolute value of its coupling to $\nu_{\mu}$ has to be much
larger than the absolute value of its coupling to $\nu_{e}$. The
above result can also be viewed as an upper bounds for $\varepsilon_{\nu_{\mu}}$.

To test above calculation, let us check how the ratio $R$ is modified
by the gauge boson X. After introducing X, the ratio is proportional
to
\begin{equation}
R\propto\bigg[\sum_{u,d}G_{F}c_{v}^{q}c_{a}^{q}\bigg]_{eff}=\sum_{u,d}\bigg[G_{F}c_{v}^{q}c_{a}^{q}+\frac{e^{2}}{\sqrt{2}}\left(\frac{\varepsilon_{l}^{\nu}c_{a}^{q}\varepsilon_{l}^{q}}{Q^{2}}\right)\bigg],\label{eq:Rnorm}
\end{equation}
where $c_{v}^{f}=I_{3}^{f}-2Q^{f}\sin^{2}\theta_{W}$, and $c_{a}^{f}=I_{3}^{f}$
are the quantum number in GWS theory. The measured value of $\bigg[\sum_{u,d}G_{F}c_{v}^{q}c_{a}^{q}\bigg]_{eff}$
is $\left(3.1507\pm0.0288\right)\times10^{-6}$, while the standard
model expectation is $3.2072\times10^{-6}$ \cite{Boehm:2004uq}.
The discrepancy can be explained by the second term in the brackets
of (\ref{eq:Rnorm}) introduced by gauge boson X. Substituting our
result for $\varepsilon_{\nu_{\mu}}$, and (\ref{eq:Feng's charges})
into this term, we find the discrepancy is indeed redeemed. Comparing
the value of $|\varepsilon_{\nu_{\mu}}|$ to the constraints in (\ref{eq:Feng constraints1}),
we notice that if the NuTeV anomaly is mainly due to the contribution
from the new vector boson X, like the quark universality, the neutrino
universality has to be broken as well.

\section{The number of neutrino flavors}

In order to check the plausibility of this $SU(3)\times SU(2)_{L}\times U(1)\times U(1)_{X}$
model, we would like to test it against to the well known number of
neutrino flavors $N_{\nu}$. This number is most precisely measured
through the $Z$ production process in $e^{+}e^{-}$collisions. The
standard model value for the ratio of the neutrino to charged leptonic
partial width is used in order to reduce the model dependence
\begin{equation}
N_{\nu}=\frac{\Gamma_{inv}}{\Gamma_{l}}\left(\frac{\Gamma_{l}}{\Gamma_{\nu}}\right)_{SM}
\end{equation}
where $\Gamma_{inv}$ is the invisible decay width of Z boson obtained
experimentally, and $\Gamma_{\nu}$ is the tree level SM expectation
of the width of Z boson decays into certain flavor of neutrino pairs.
$\Gamma_{\nu}$ represents the invisible partial width, which is determined
by subtracting the visible partial widths from the total Z width.
It is assumed that each light neutrino flavor makes the identical
contribution $\Gamma_{\nu}$ to the neutrino partial width due to
lepton universality. The visible width corresponds to Z decays into
quarks and charged leptons. A combination of several experimental
measurements gives the result $N_{\nu}=2.984\pm0.008$ \cite{Agashe:2014kda}.
To find out if the propagator of X boson will alter $N_{\nu}$ significantly,
we calculate the distribution of the cross-section for the process
$e^{+}e^{-}\to\nu\bar{\nu}$ shown in FIG.1. Since lepton universality
is broken in the extended model, the contribution of the decay to
each neutrino flavor is calculated separately. In our computation,
we take the upper bounds of the coupling constants, and assume X boson
equally couples to muon neutrino and tau neutrino.
\begin{center}
\includegraphics[width=7cm]{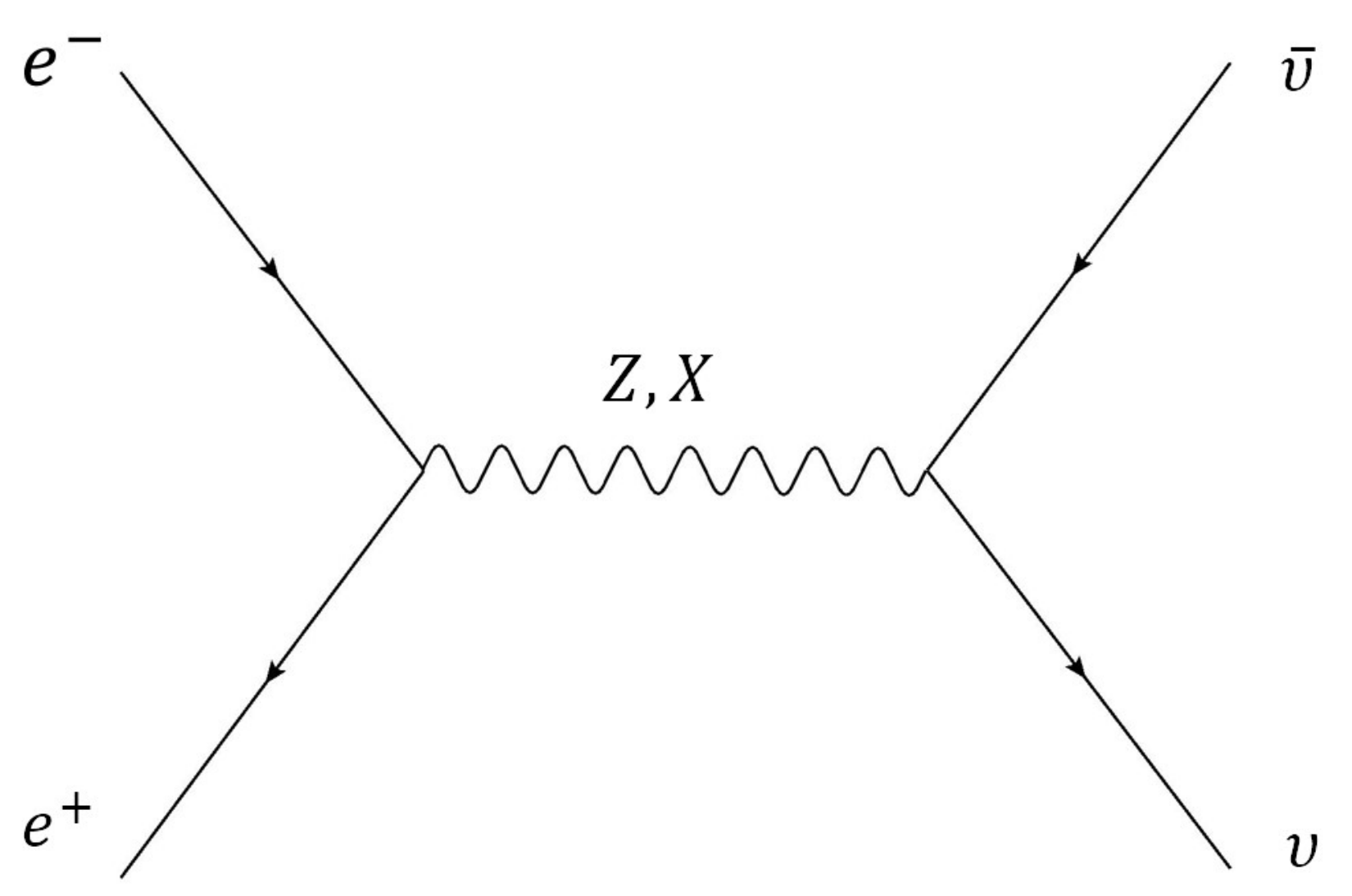}
\figcaption{\label{fig1}   The leading diagrams that contribute to the X-boson production in
electron-positron collision. }
\end{center}

The result is displayed in FIG.2, where we can see that even the upper
bounds of the coupling constants is too small to arise any noticeable
effect on the decay width of the Z boson. Our result for the coupling
constant $\varepsilon_{\nu_{\mu}}$ is safe from being contradicted
to the well tested conclusion of the number of neutrino flavors.
\begin{center}
\includegraphics[width=7cm]{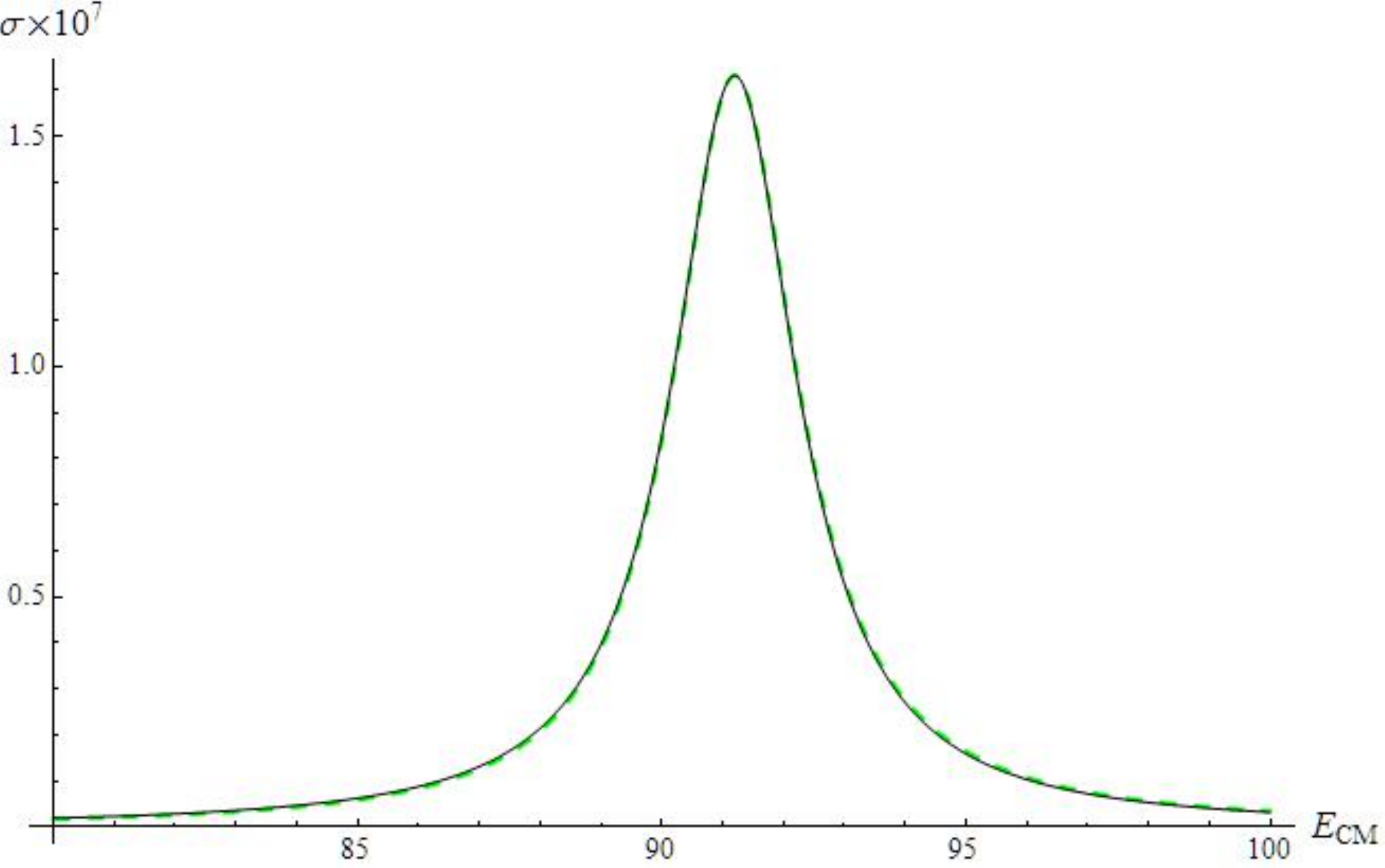}
\figcaption{\label{fig2}   Taking the upper bounds of $\varepsilon_{e}$, and $|\varepsilon_{\nu_{\mu}}|=6\times10^{-3}$
from previous subsection, we calculate the distribution of the cross-section
for the process $e^{+}e^{-}\to\nu\bar{\nu}$. The solid line is the
SM prediction, the green dashed line is the distribution. }
\end{center}

\section{Neutrino Trident Production}

Models based on gauged muon number $L_{\mu}$ is strictly constrained
by the SM trident production of neutrino, where a pair of muon and
anti-muon is produced in the scattering of muon neutrinos in the Coulomb
field of a target nucleus. New force mediated by a heavy vector boson
is excluded as a solution of the $\left(g-2\right)_{\mu}$ anomaly
\cite{Altmannshofer:2014cfa}. The existence of a vector gauge boson
with a mass of $16.7\:\mathrm{MeV}$ would also be excluded in the
$L_{\mu}-L_{\tau}$ model \cite{Altmannshofer:2014pba} by combining
the measurements of the neutrino trident production with our result
of $\varepsilon_{\nu_{\mu}}$. We will show here that there is room
for X boson, if the simple $SU(3)\times SU(2)_{L}\times U(1)\times U(1)_{X}$
model is adopted.

The contribution of X to trident production of neutrino at tree level
is shown in FIG.3. In SM, the propagator of X is replaced with the
W and Z boson propagator. Unlike in the $L_{\mu}-L_{\tau}$ model,
the X couplings to muons and muon-neutrinos may have the same or opposite
sign. Therefore, the trident production may be reduced or enhanced
by the interference between the X-boson and W-boson (Z-boson). We
adopt a calculation procedure using the equivalent photon approximation
(EPA) \cite{Altmannshofer:2014pba,vonWeizsacker:1934nji,Williams:1934ad}.
The full cross-section of a neutrino scattering with a nucleus N can
be written as a convolution of two separate parts
\begin{equation}
\sigma\left(\nu_{\mu}N\to\nu_{\mu}N\mu^{+}\mu^{-}\right)=\int\sigma\left(\nu_{\mu}\gamma\to\nu_{\mu}\mu^{+}\mu^{-}\right)P\left(s,q^{2}\right),
\end{equation}
where the first part of the integrand $\sigma\left(\nu_{\mu}\gamma\to\nu_{\mu}\mu^{+}\mu^{-}\right)$
is the cross-section for a neutrino scattered off a real photon; the
second part $P\left(s,q^{2}\right)=\mbox{\ensuremath{\frac{Z^{2}e^{2}}{4\pi^{2}}\frac{ds}{s}\frac{dq^{2}}{q^{2}}}}F^{2}\left(q^{2}\right),$
is the probability of creating a virtual photon with virtuality $q^{2}$
and energy $\sqrt{s}$ in the center-of-mass frame of the neutrino
and a real photon. The virtual photon is created in the electromagnetic
field of the nucleus N with charge $Ze$ and a electromagnetic form-factor
(FF) $F\left(q^{2}\right)$. Generally, the real photon cross-section
can be written as
\begin{equation}
\sigma^{\left(\mathrm{SM+X}\right)}=\sigma^{\left(\mathrm{SM}\right)}+\sigma^{\left(\mathrm{inter}\right)}+\sigma^{\left(\mathrm{X}\right)},
\end{equation}
where the second term comes from the interference between the SM and
the X contributions. The differential cross-sections for each of them
have a general symbolical form
\begin{equation}
d\sigma=\frac{1}{2s}\mathrm{dPS_{3}}\left(\frac{1}{2}M^{2}\right)\frac{G_{F}^{2}e^{2}}{2}
\end{equation}
Here, $G_{F}=\sqrt{2}g^{2}/\left(8M_{W}^{2}\right)$ is the Fermi
constant, and $\mathrm{dPS_{3}}$ is the 3-body phase-space. In our
calculation, the squared amplitudes $M^{2}$ are generated by FeynCalc
\cite{Shtabovenko:2016olh}. By replacing the propagator with one
over mass of the mediator boson squared, and omitting terms proportional
to the muon mass in the numerator, we recover the SM expression given
in \cite{Belusevic:1987cw,Altmannshofer:2014pba}. The phase-space
integration is numerically calculated with Vegas \cite{Hahn:2005pf}.
Our calculation verified the analytic expression of the leading log
approximation for real photon cross-section in SM \cite{Altmannshofer:2014pba}.

By numerically integrating the real-photon cross-section with the
probability distribution function $P\left(s,q^{2}\right)$ in the
range of $4m_{\mu}^{2}<s<2E_{\nu_{\mu}}q$ and $2m_{\mu}^{2}/E_{\nu_{\mu}}<q<\infty$,
we obtain the total cross-section for $\nu_{\mu}N\to\nu_{\mu}N\mu^{+}\mu^{-}$.
We use a simple exponential function to mimic the nucleus form factor
\cite{Lovseth:1971vv}. In order to test our calculation, we reproduced
the prediction of SM and V-A theory \cite{Lovseth:1971vv,Brown:1973ih}.

Nuetrino trident production has been studied by several experiments
\cite{Geiregat:1990gz,Mishra:1991bv,Adams:1999mn}, among which, the
measurement from CCFR collaboration provide the strongest constrains
on the parameter space, and is used in our study. The CCFR collaboration
detected the trident events by scattering a neutrino beam with mean
energy of $\mbox{E=160}\:\mathrm{GeV}$ with an iron target. The ratio
of the cross-section they obtained to the SM prediction is $\sigma_{CCFR}/\sigma_{SM}=0.82\pm0.28$.
At this energy level, it is more secured not to take any approximation
in the formulation of the amplitudes. In our calculation, we keep
all the gauge boson propagators, and all the terms containing muon
mass. By combining CCFR measurement with our numerical result, we
obtain the following range for the first-generation charge of the
gauge boson X
\begin{equation}
-2.0\times10^{-5}<\varepsilon_{\nu_{\mu}}\varepsilon_{\mu}<6\times10^{-7}
\end{equation}
We notice here that if $\varepsilon_{\nu_{\mu}}$ and $\varepsilon_{\mu}$
have the same sign, and particle X is fully responsible for the NuTeV
anomaly, the value of $\varepsilon_{\mu}$ is strictly restricted
to be less than $3\times10^{-4}$, which exclude the possibility for
the gauge boson X to be the solution of the $\left(g-2\right)_{\mu}$
anomaly \cite{Pospelov:2008zw}. However, if $\varepsilon_{\nu_{\mu}}$
and $\varepsilon_{\mu}$ have opposite signs, the constrain on $\varepsilon_{\mu}$
is greatly relaxed to $|\varepsilon_{\mu}|<1\times10^{-2}$, making
it a candidate for solving $\left(g-2\right)_{\mu}$ puzzle. Future
experiment such as LBNE may provide more data on neutrino trident
production \cite{Altmannshofer:2014pba}, which may lead to decisive
analysis on the coupling of X(16.7) to neutrinos.
\begin{center}
\includegraphics[width=7cm]{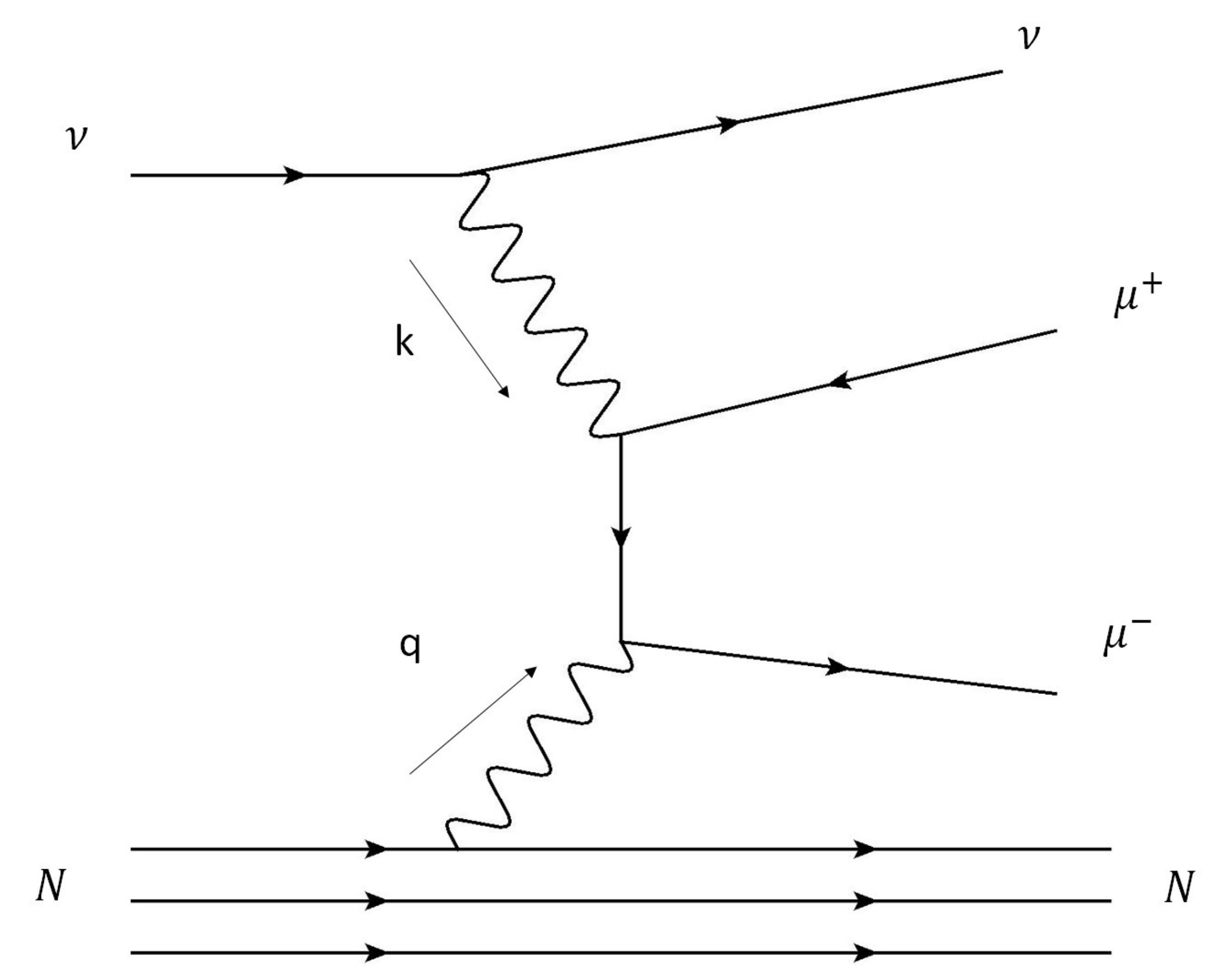}
\figcaption{\label{fig3}   {The trident process at tree level} }
\end{center}

\section{Conclusions}

Unlike heavy $Z'$ boson that has been massively discussed in the
literature \cite{Gauld:2013qba,Buras:2013qja,Gauld:2013qja,Buras:2013dea},
the newly found gauge boson X is very light. It is quite exciting
to know that low energy experiments still allow a possibility for
finding such a light boson. A commonly asked question is what are
the constrains for this new particle from preexisting experimental
measurements. We investigate some of the consequences brought by this
unusual vector gauge boson X. The SM gauge group $SU(3)\times SU(2)_{L}\times U(1)$
extended by an Abelian gauge groups $U(1)_{X}$ was adopted in our
calculation. Its implication on the NuTeV anomaly was studied. First
of all, we found that the charge has to be $|\varepsilon_{\nu_{\mu}}|\simeq2\times10^{-3}$,
with the opposite sign to $\varepsilon_{u}$, in order to attribute
the NuTeV anomaly entirely to the gauge boson X(16.7). We have proven
that this value, although is comparable or even larger than the coupling
constants for the other fermions, is still too small to allow any
effect on the experimental measurement of the number of neutrino flavors
to be noticed in previous experiments. Next, we studied the neutrino
trident production. Comparing the numerical calculation of this process
with measurements from CCFR results in a powerful constraint on the
parameter space of the model: $-2.0\times10^{-5}<\varepsilon_{\nu_{\mu}}\varepsilon_{\mu}<6\times10^{-7}$.
When combined with the requirement of explaining the discrepancy in
the muon $\left(g-2\right)$, unlike what would happen to a heavy
vector boson \cite{Altmannshofer:2014cfa}, the light gauge boson
X(16.7) survived. Particularly, if $\varepsilon_{\nu_{\mu}}$ and
$\varepsilon_{\mu}$ have the same sign, the vector gauge boson X
cannot be responsible for both the NuTeV and the $\left(g-2\right)_{\mu}$
anomaly. However, if $\varepsilon_{\nu_{\mu}}$ and $\varepsilon_{\mu}$
have opposite signs, X(16.7) can indeed be the solution to both of
these puzzles. On the other hand, $|\varepsilon_{\nu_{\mu}}|$ would
be smaller, if other effect such as the strange sea asymmetry or isospin
violation take partial responsibility for the discrepancy between
NuTeV and the SM prediction. In that case, a gauge boson X with $\varepsilon_{\nu_{\mu}}\varepsilon_{\mu}>0$
can be the solution of the $\left(g-2\right)_{\mu}$ anomaly.
Finally, although the coupling of X boson to muon neutrinos deduced from NuTeV anomaly
is significantly larger than the coupling of X to electron neutrino
\cite{Feng:2016jff}, it survives the constraint deduced from the CHARM II experiment \cite{Bilmis:2015lja}, if the uncertainties of measurements are taken into account. This value of coupling may lead to a deformation of the invariant
mass distribution of $e^{+}e^{-}$ in final state for the differential
cross section proposed to search X boson \cite{Chen:2016dhm}.

\acknowledgments{We appreciate very much the discussion with Dr. Chang-Zheng Yuan and Dr. Jing-Zhi
Zhang on detecting X at BESIII. We thank Dr. Jun Jiang for many helpful comments and discussions regarding this work.}

\end{multicols}

\vspace{8mm}

\centerline{\rule{80mm}{0.1pt}}
\vspace{2mm}

\begin{multicols}{2}

\end{multicols}

\clearpage
\end{document}